\documentclass[letterpaper,12pt]{article}
\usepackage[nodisplayskipstretch]{setspace}
\usepackage[stretch=10,shrink=10]{microtype}
\usepackage{titlesec}
\titlespacing*{\section}{0pt}{2.0ex plus .2ex minus .2ex}{1.2ex plus .2ex minus .2ex}
\titlespacing*{\subsection}{0pt}{2.0ex plus .2ex minus .2ex}{1.2ex plus .2ex minus .2ex}
\usepackage[T1]{fontenc}
\usepackage[utf8]{inputenc}

\usepackage{amsmath,amssymb,amsthm,mathtools}
\usepackage{enumitem}
\usepackage{url}
\usepackage[letterpaper,margin=1in]{geometry}
\usepackage{xcolor}
\usepackage{graphicx}
\usepackage{booktabs,longtable,array,tabularx}
\usepackage{tikz}
\usetikzlibrary{arrows.meta,positioning,calc,patterns,decorations.pathreplacing,shapes.geometric,fit}
\usepackage{pgfplots}
\pgfplotsset{compat=1.18}
\usepackage{hyperref}
\hypersetup{colorlinks=true,linkcolor={red!80!black},citecolor={black!50!blue},filecolor=magenta,urlcolor=blue}
\usepackage{csquotes}
\usepackage[backend=biber,style=apa]{biblatex}
\usepackage[capitalise,noabbrev]{cleveref}
\usepackage{newtxtext}

\setlist[enumerate]{itemsep=0pt,topsep=3pt}
\allowdisplaybreaks[2]

\theoremstyle{plain}
\newtheorem{theorem}{Theorem}
\newtheorem{proposition}{Proposition}
\newtheorem{lemma}{Lemma}

\theoremstyle{definition}
\newtheorem{definition}{Definition}

\theoremstyle{plain}

\numberwithin{equation}{section}
\crefname{fieldconditions}{Conditions}{Conditions}
\Crefname{fieldconditions}{Conditions}{Conditions}

\newcommand{\R}{\mathbb{R}}
\newcommand{\one}{\mathbf{1}}
\newcommand{\eps}{\varepsilon}
\newcommand{\dd}{\,\mathrm{d}}
\newcommand{\Hone}{\mathcal{H}^1}
\newcommand{\Leb}{\mathcal{L}^2}
\DeclareMathOperator{\dv}{div}

\newcommand{\cst}{c_{*}}

\title{Revenue-optimal auctions for two additive bidders with two uniformly valued items}
\author{%
  Genta Okada\thanks{Corresponding author. Graduate School of Economics, The University of Tokyo, 7-3-1 Hongo, Bunkyo-ku, Tokyo, 113-0033, Japan. E-mail: \href{mailto:ogenta0628@g.ecc.u-tokyo.ac.jp}{\nolinkurl{ogenta0628@g.ecc.u-tokyo.ac.jp}}.}\qquad
  Sho Mikami\thanks{Graduate School of Economics, The University of Tokyo, 7-3-1 Hongo, Bunkyo-ku, Tokyo, 113-0033, Japan. E-mail: \href{mailto:mikami-sho910@g.ecc.u-tokyo.ac.jp}{\nolinkurl{mikami-sho910@g.ecc.u-tokyo.ac.jp}}.}
}
\date{November 2026}

\begin{document}
\maketitle

\begin{abstract}
We study the revenue-maximization problem with two bidders who have additive valuations for two heterogeneous items. When all four valuations are independently drawn from the uniform distribution on $[c,c+1]$, we show that a simple deterministic auction belonging to the class of affine maximizers is optimal for sufficiently large $c$. We further establish a similar optimality result for every $c\geq0$ when the seller is required to allocate both items. Both results hold among all dominant-strategy incentive compatible and ex-post individually rational mechanisms, including randomized mechanisms. This is the first closed-form solution to the revenue-maximization problem with multiple additive bidders, multiple items, and i.i.d. continuous item valuations.


\end{abstract}

\section{Introduction}\label{sec:intro}
\label{sec:problem}

Consider a seller who owns two heterogeneous items and faces two buyers.
Each buyer privately knows her valuation for each item and values the bundle
at the sum of these valuations. The seller knows the distribution of
valuations and seeks to maximize expected revenue. Should the seller auction
the items separately, sell them together as a bundle, or combine the two
selling methods?

For a single item, \textcite{Myerson1981} shows how the seller can maximize
revenue by comparing the buyers' virtual valuations. With multiple items,
however, determining which buyer receives one item can affect the payments
that the seller can extract for other items. There are two natural ways to
exploit competition between buyers. Separate selling makes the buyers
compete for each item, whereas pure bundling makes them compete for the
bundle. Neither method is generally optimal. \textcite{Palfrey1983}
initiates the comparison of bundling and separate auctions, and
\textcite{JMM2007} show that mixed-bundling auctions can generate strictly
higher revenue than either pure bundling or separate efficient auctions.

These comparisons suggest a further question: when is a mixed-bundling
auction optimal among all incentive-compatible mechanisms? A mechanism may
condition the allocation and payment for one item on reports about the
other item, use lotteries, or treat buyers asymmetrically. Moreover, the
menus offered to different buyers must be jointly feasible at every
report profile. Optimizing over a family of simple auctions does not
establish that these additional possibilities are unprofitable.

We address this question in a symmetric environment with two buyers and two
items. The four valuations are independently drawn from the uniform
distribution on $[c,c+1]$, where $c\ge0$. Buyers have additive valuations and
quasi-linear utility. The seller maximizes expected revenue subject to
dominant-strategy incentive compatibility (DSIC) and ex-post individual
rationality (IR). In this environment, increasing $c$ raises the value of
every item without changing the dispersion of valuations or the differences
between buyers' valuations. This allows us to distinguish the problem of
allocating the items among buyers from the problem of deciding whether to
sell them. When $c$ is large, withholding an item is costly to the seller,
but even a small possibility of withholding may help reduce buyers'
information rents.

Our first result characterizes an optimal auction under complete sale:
both items must be allocated at almost every type profile. For every
$c\ge0$, a deterministic mixed-bundling auction is optimal. Up to ties,
it splits the items when each buyer values a different item more than
the other buyer does by more than $1/3$; otherwise, the buyer with the
higher total valuation receives both. The auction is an affine maximizer
with a penalty of $1/3$ on split allocations, and its expected revenue is
\[
2c+\frac{637}{810}.
\]
The split penalty and revenue at $c=0$ were obtained by
\textcite[Example~3]{JMM2007} through optimization within a parametric
family. We establish optimality over the entire class of DSIC,
ex-post IR mechanisms satisfying complete sale. The result extends to
every shift because complete sale fixes the contribution of $c$ to
expected revenue at $2c$.

The split penalty creates a discount for buying both items together.
Each item costs the other buyer's valuation for that item plus $1/3$,
whereas the bundle costs the other buyer's total valuation. The auction
thus trades off item-level competition against bundle-level competition:
it favors bundling even at the expense of allocative efficiency, unless
the buyers' comparative advantages are sufficiently large.

Our second result establishes optimality without the complete-sale
restriction. For every $c\ge91/15$, it suffices to add a bundle reserve
to the same mixed-bundling auction. The reserve is $2c+\eps(c)$, where
\[
\eps(c)=\frac{\sqrt{4c^2+6}-2c}{3}.
\]
If both buyers' total valuations are at most this reserve, neither item
is sold. Otherwise, the allocation coincides with that of the no-reserve
auction. The mechanism is again an affine maximizer, now with a fifth
outcome corresponding to no sale. Its optimal expected revenue is
\[
2c+\frac{637}{810}+\frac{\eps(c)^3}{12}
+\frac{3\eps(c)^5}{40}.
\]
Neither result restricts the seller to deterministic, symmetric, or
affine-maximizing mechanisms. The threshold $91/15$ is sufficient for
the unrestricted problem; we do not claim that it is sharp.

Two features of the result are worth emphasizing. First, the optimal split
penalty does not depend on the shift, while the reserve premium $\eps(c)$
does. In particular, allowing the seller to withhold items does not require
changing the allocation rule outside the no-sale region. Second, the
bundle reserve coincides with the optimal pure-bundling price for a single
buyer with two independent $U[c,c+1]$ valuations in the large-shift regime
studied by \textcite{Pavlov2011}. Competition changes how items are allocated
when they are sold, but the optimal reserve in our environment retains
this single-buyer form. Even for arbitrarily large $c$, the reserve strictly
improves expected revenue over the optimal complete-sale auction.
The improvement is asymptotic to $1/(96c^3)$. Thus complete sale is
a useful intermediate problem, not a restriction that can simply be imposed
without loss for large valuations.

The main difficulty is to establish optimality over the entire
admissible class. We derive a dual optimality criterion that separates
pointwise virtual-welfare maximization from a conditional utility
bound. A common dual field assigns the same virtual valuations to
both buyers on bundle outcomes, while two affine fields favor the
specified recipients on split outcomes.

The conditional utility bound relies on individual rationality and
only a subset of the incentive constraints. Under complete sale,
coordinatewise incentive constraints suffice. With a reserve, the
proof additionally uses monotonicity of the difference in allocation
probabilities along certain equal-total-value segments. Neither
argument requires joint convexity of the bidder's utility in her two
valuations. The same criterion thus establishes both optimality
results, with an additional field correction for the reserve problem.

\paragraph{Related literature.}\label{sec:lit}
Our closest antecedents are the studies of mixed bundling with competing
buyers. \textcite{JehielMoldovanu2001} establish revenue gains from bundling
relative to efficient auctions in a two-buyer environment with a
complete-sale requirement. \textcite{JMM2007} extend this idea by allowing
the auction to favor particular partitions of the items among buyers.
\textcite{TangSandholm2012} incorporate reserve prices and optimize revenue
within a class of mixed-bundling auctions. We complement these studies by showing that, under complete sale
and in the unrestricted large-shift environment, the simple
mixed-bundling form is optimal without restricting the seller to a
parametric family.

Our model is also related to the literature on optimal selling mechanisms
for a single buyer. \textcite{ManelliVincent2006} characterize the
optimality of particular deterministic menus, and \textcite{Pavlov2011}
solves the two-item problem in shifted-uniform examples. Using a dual
formulation in terms of signed measures and optimal transport,
\textcite{DDT2017} extend the large-shift pure-bundling result to an
arbitrary fixed number of items. \textcite{GK2014} develop a dual framework
and apply it to the single-buyer uniform problem. With two competing
buyers, the seller can benefit from splitting the items rather than
always selling a single bundle, and feasibility couples the menus of
the buyers. Our result characterizes the optimal way to combine
bundling, splitting, and reserve pricing in this setting.

Several papers obtain exact solutions to multi-buyer, multi-item
revenue-maximization problems under other assumptions on types or
incentive constraints. \textcite{Levin1997} and \textcite{Ulku2013}
study one-dimensional signals in combinatorial auctions.
\textcite{Armstrong2000} and \textcite{TangWang2016} obtain results for
Bayesian incentive-compatible auctions, including continuous examples
with restricted type supports. For independent additive valuations,
\textcite{Yao2017} derives exact DSIC and Bayesian optimal revenues
when each valuation has a two-point support.
Related recent work studies optimal mechanisms or the optimality
of separate selling in finite-support environments
\parencite{MertzanidisTerzoglou2025,LiuLiuWang2026,Mikami2026}.
In contrast, our buyers each have a two-dimensional continuous type
with independent coordinates, and our result gives an explicit
revenue-maximizing DSIC mechanism in this environment.

Methodologically, our proof builds on dual approaches to
multidimensional mechanism design.
\textcite{CDW2012} and \textcite{CDW2016} relate Bayesian revenue
maximization to virtual-welfare maximization in finite environments,
while \textcite{Zuo2017} develops a dual formulation for DSIC auctions.
\textcite{KSTZ2022} establish strong duality for continuous multi-buyer,
multi-item Bayesian auctions. Related computational approaches
construct candidate auctions and upper bounds on revenue
\parencite{CurrySandholmDickerson2023,DuettingEtAl2024,WangJiangParkes2024,JiangParkesWang2026}.
Our contribution is an explicit continuous-domain dual certificate
that is tight at a deterministic auction and whose verification
requires only the restricted incentive inequalities described above.

The rest of the paper proceeds as follows.
\Cref{sec:model} presents the auction and main results.
\Cref{sec:field} establishes the dual optimality criterion, and
\cref{sec:corner} applies it to prove both results.
\Cref{sec:remarks} concludes. \Cref{app:proofs} derives the auction's
properties and revenue formulas, and the remaining appendices
construct and verify the dual fields.

The Main Theorem, including the formula for the revenue, has also been verified formally in the proof assistant Lean~4; the code is available at \url{https://github.com/silsilatun/optimal_auction_lean}.

\section{Model and main results}\label{sec:model}

We first specify the model and admissible mechanisms, then describe the candidate auction and state \cref{thm:main}.

\subsection{Mechanisms}\label{sec:mechanisms}

There are two risk-neutral bidders and two heterogeneous items. Bidder~$i$ values item~$j$ at $v_{ij}$, where $i,j\in\{1,2\}$. All four valuations are independently drawn from $U[c,c+1]$, with a common shift $c\ge0$. Write $v_i=(v_{i1},v_{i2})$ for bidder~$i$'s valuation vector. Values are additive and utilities are quasi-linear.

Each bidder's type space is $X_c=[c,c+1]^2$. Write $v=(v_1,v_2)\in X_c\times X_c$ for the type profile and $\hat v=(\hat v_1,\hat v_2)\in X_c\times X_c$ for a report profile. The notation $(\hat v_i,v_{-i})$ means the profile obtained from $v$ by replacing bidder~$i$'s component by $\hat v_i$, with the bidders kept in their original order. Write $\one=(1,1)$, so that $\one\cdot v_i$ is bidder~$i$'s valuation for the bundle. Vector inequalities are componentwise.

\begin{definition}[Direct mechanism]\label{def:mech}
A direct mechanism $M=(q,p)$ consists of an allocation rule $q$ and a payment rule $p$, with Borel measurable components
\[
q_i:X_c\times X_c\to[0,1]^2,\qquad
p_i:X_c\times X_c\to\R,\qquad i=1,2,
\]
where the payment functions are integrable. Given reports $\hat v$, bidder~$i$ receives item~$j$ with probability $q_{ij}(\hat v)$ and makes expected payment $p_i(\hat v)$. Write $q_i=(q_{i1},q_{i2})$ for bidder~$i$'s allocation vector. Feasibility requires
\[
\sum_{i=1}^2 q_{ij}(\hat v)\le1
\qquad(j=1,2,\ \hat v\in X_c\times X_c).
\]
A bidder with true type $v_i$ obtains utility $v_i\cdot q_i(\hat v)-p_i(\hat v)$ at report profile $\hat v$. In particular, bidder~$i$'s truthful utility $u_i:X_c\times X_c\to\R$ is
\[
u_i(v)=v_i\cdot q_i(v)-p_i(v).
\]
The mechanism is must-sell if $q_1+q_2=\one$ almost everywhere. Its expected revenue under truthful reporting is
\[
\mathrm{Rev}(M)=\mathbb E_v\left[\sum_{i=1}^2p_i(v)\right].
\]
Allocations, payments and utilities are evaluated in expectation over the mechanism's own randomization, conditional on the reports.
\end{definition}

The bidders are risk neutral and their values are additive, so only the probabilities $q_{ij}$ and the expected payments matter; a mechanism that uses a lottery is described completely by these expectations.
Conversely, any $q_1,q_2$ with $q_1+q_2\le\one$ can be realized by a lottery: give item $j$ to bidder~1 with probability $q_{1j}$, to bidder~2 with probability $q_{2j}$, and keep it otherwise.
So \cref{def:mech} contains all randomized mechanisms.
Nothing in the definition is symmetric in the two bidders, and we never assume symmetry.

\begin{definition}[DSIC, ex-post IR]\label{def:dsic}
A mechanism is dominant-strategy incentive compatible (DSIC) if
\[
u_i(v_i,v_{-i})\ge
u_i(\hat v_i,v_{-i})+(v_i-\hat v_i)\cdot q_i(\hat v_i,v_{-i})
\]
for every $i=1,2$, $v\in X_c\times X_c$ and $\hat v_i\in X_c$. Thus truthful reporting is a best response to every report of the other bidder.

A mechanism is ex-post individually rational (IR) if $u_i(v)\ge0$ for every $i=1,2$ and $v\in X_c\times X_c$.
\end{definition}

\subsection{The candidate auction}\label{sec:mech}

The candidate is a mixed-bundling auction that penalizes split allocations by $d=1/3$ and imposes a reserve price of $2c+\eps$ on the bundle, where
\begin{equation}\label{eq:eps}
\eps=\eps(c)=\frac{\sqrt{4c^2+6}-2c}{3}.
\end{equation}
Equivalently, $\eps$ is the positive root of $3\eps^2+4c\eps=2$. The reserve coincides with the optimal pure-bundling price for a single buyer in this environment \parencite[Example~3]{Pavlov2011}; its derivation is given in \cref{app:proofs-model}. In particular, $\eps(c)\sim 1/(2c)$ as $c\to\infty$, and $\eps(c)\le d$ if and only if $c\ge5/4$.

At each report profile $\hat v$, the auction selects the allocation that maximizes the adjusted welfare in the following table.
\begin{center}
\begin{tabular}{lll}
\toprule
Outcome & Allocation & Adjusted welfare\\
\midrule
$\mathrm B_1$ & Both items to bidder 1 & $\hat v_{11}+\hat v_{12}$\\
$\mathrm B_2$ & Both items to bidder 2 & $\hat v_{21}+\hat v_{22}$\\
$\mathrm S_{12}$ & Item 1 to bidder 1, item 2 to bidder 2 & $\hat v_{11}+\hat v_{22}-d$\\
$\mathrm S_{21}$ & Item 1 to bidder 2, item 2 to bidder 1 & $\hat v_{12}+\hat v_{21}-d$\\
$\mathrm N$ & No sale & $2c+\eps$\\
\bottomrule
\end{tabular}
\end{center}
Ties are resolved in the fixed priority order $\mathrm N,\mathrm B_1,\mathrm B_2,\mathrm S_{12},\mathrm S_{21}$.

Payments are the corresponding Clarke-type payments. Equivalently, fixing the other bidder's report $\hat v_{-i}$, bidder~$i$ faces the following menu:
\begin{center}
\begin{tabular}{ll}
\toprule
Bundle & Price\\
\midrule
Nothing & $0$\\
Item $j$ ($j=1,2$) & $\max\{\one\cdot \hat v_{-i},2c+\eps\}-\hat v_{-i,3-j}+d$\\
Both items & $\max\{\one\cdot \hat v_{-i},2c+\eps\}$\\
\bottomrule
\end{tabular}
\end{center}

Denote this auction by $M_\eps=(q^\eps,p^\eps)$ and its counterpart without the no-sale outcome by $M_0=(q^0,p^0)$. The menu of $M_0$ is obtained by replacing $\max\{\one\cdot \hat v_{-i},2c+\eps\}$ with $\one\cdot \hat v_{-i}$. Both mechanisms also depend on $c$, which is suppressed in the notation. Superscripts $0$ and $\eps$ on allocations, payments, and utilities indicate the corresponding mechanism.

Both auctions are affine maximizers with outcome-specific constants \parencite{JMM2007,CurrySandholmDickerson2023}. They are DSIC and ex-post IR under the specified payments, and the single maximizing outcome selected by the priority rule ensures joint feasibility. The payment formula and incentive properties are verified in \cref{app:proofs-model}.

In $M_0$, the split penalty induces a discount for buying both items together: each individual item costs the opponent's valuation for that item plus $d$, whereas the bundle costs the opponent's total valuation. Up to ties, the auction splits the items when the two comparative advantages both exceed $d$, and otherwise awards the bundle to the bidder with the higher total valuation. For $c\ge5/4$, adding the reserve changes this allocation only when both bidders' totals are at most $2c+\eps$, in which case neither item is sold.

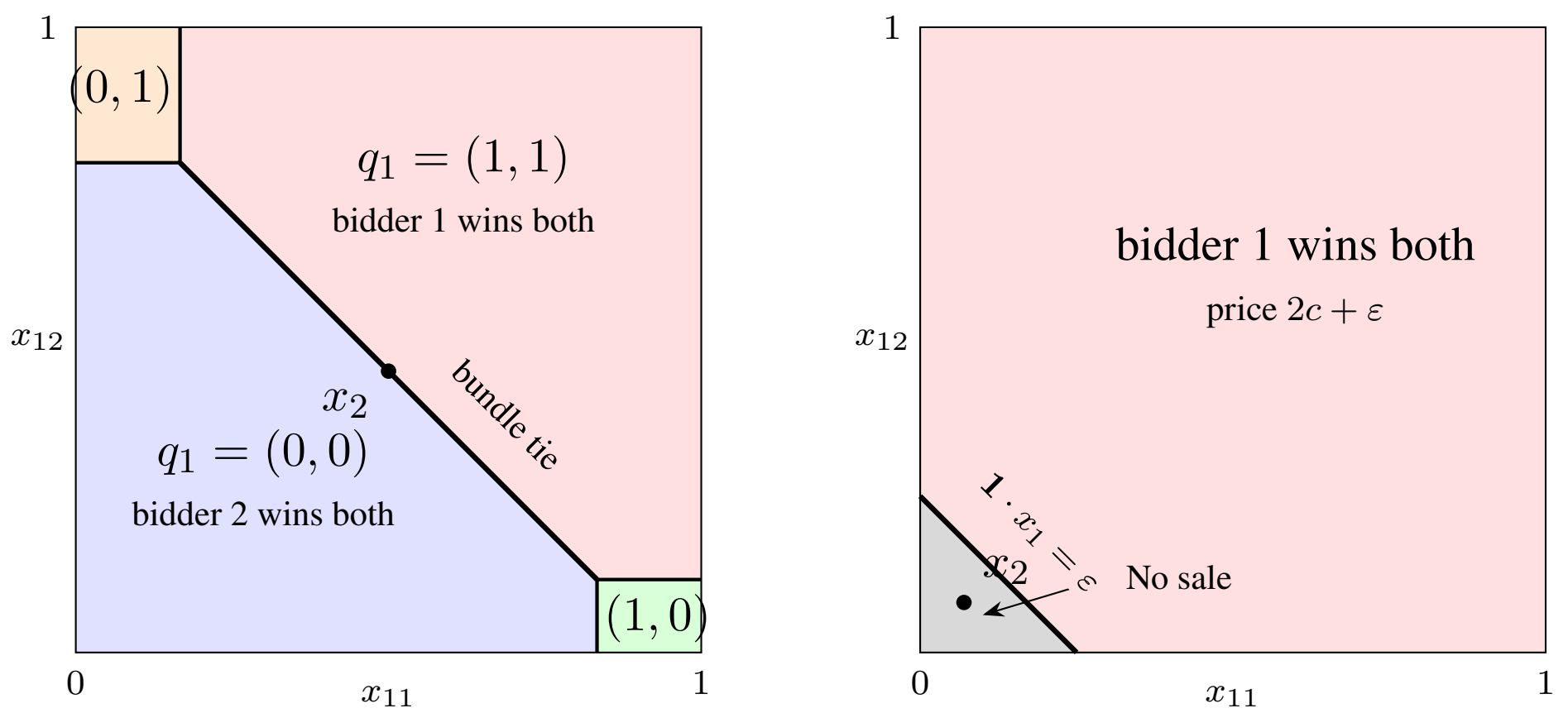
\begin{figure}[!htbp]
\centering
\begin{tikzpicture}[scale=5.0,>=Stealth]
\begin{scope}
\def\za{0.5}\def\zb{0.45}
\fill[red!12] (1/6,47/60)--(1/6,1)--(1,1)--(1,7/60)--(5/6,7/60)--cycle;
\fill[blue!12] (0,0)--(5/6,0)--(5/6,7/60)--(1/6,47/60)--(0,47/60)--cycle;
\fill[green!15] (5/6,0)--(1,0)--(1,7/60)--(5/6,7/60)--cycle;
\fill[orange!18] (0,47/60)--(1/6,47/60)--(1/6,1)--(0,1)--cycle;
\draw (0,0) rectangle (1,1);
\draw[very thick] (1/6,47/60)--(5/6,7/60);
\draw[thick] (1/6,47/60)--(1/6,1) (5/6,0)--(5/6,7/60)--(1,7/60) (0,47/60)--(1/6,47/60);
\fill (\za,\zb) circle (0.35pt) node[below left,font=\small]{$x_2$};
\node[font=\small,align=center] at (0.62,0.75) {$q_1=(1,1)$\\ {\scriptsize bidder 1 wins both}};
\node[font=\small,align=center] at (0.3,0.28) {$q_1=(0,0)$\\ {\scriptsize bidder 2 wins both}};
\node[font=\small] at (0.93,0.055) {$(1,0)$};
\node[font=\small] at (0.07,0.9) {$(0,1)$};
\node[font=\scriptsize,rotate=-45,fill=red!12,inner sep=1pt] at (0.69,0.38) {bundle tie};

\node[below,font=\scriptsize] at (0,0) {$0$};
\node[below,font=\scriptsize] at (1,0) {$1$};
\node[left,font=\scriptsize] at (0,1) {$1$};
\node[font=\scriptsize] at (0.5,-0.07) {$x_{11}$};
\node[font=\scriptsize] at (-0.06,0.5) {$x_{12}$};
\node[font=\small] at (0.5,-0.15) {(a) $\one\cdot x_2>\eps$, here $x_2=(\frac12,\frac9{20})$};
\end{scope}
\begin{scope}[xshift=1.35cm]
\fill[red!12] (0.25,0)--(1,0)--(1,1)--(0,1)--(0,0.25)--cycle;
\fill[gray!30] (0,0)--(0.25,0)--(0,0.25)--cycle;
\draw (0,0) rectangle (1,1);
\draw[very thick] (0.25,0)--(0,0.25);
\fill (0.07,0.08) circle (0.35pt) node[above right,font=\small]{$x_2$};
\node[font=\small,align=center] at (0.6,0.6) {bidder 1 wins both\\ {\scriptsize price $2c+\eps$}};
\node[font=\scriptsize,align=center,anchor=west] at (0.3,0.12) {No sale};
\draw[->,thin] (0.3,0.12)--(0.1,0.06);
\node[font=\scriptsize,rotate=-45,fill=red!12,inner sep=1pt] at (0.19,0.19) {$\one\cdot x_1=\eps$};
\node[below,font=\scriptsize] at (0,0) {$0$};
\node[below,font=\scriptsize] at (1,0) {$1$};
\node[left,font=\scriptsize] at (0,1) {$1$};
\node[font=\scriptsize] at (0.5,-0.07) {$x_{11}$};
\node[font=\scriptsize] at (-0.06,0.5) {$x_{12}$};
\node[font=\small] at (0.5,-0.15) {(b) $\one\cdot x_2<\eps$ ($\eps$ enlarged)};
\end{scope}
\end{tikzpicture}
\caption{Allocation of $M_\eps$ in bidder~1's translated type square for a fixed opponent $x_2$. (a) Above the reserve, the allocation coincides with $M_0$; the labels give bidder~1's allocation vector. (b) Below the reserve, bidder~1 buys the bundle at price $2c+\eps$ if $\one\cdot x_1>\eps$; otherwise nothing is sold (gray triangle). The reserve premium in panel (b) is enlarged for illustration.}
\label{fig:slice}
\end{figure}

To illustrate the allocation rule, translate each bidder's type to
$x_i=v_i-c\one\in X=[0,1]^2$. The translated types are independent
and uniform on $X$, and $v_1-v_2=x_1-x_2$. Write $x=(x_1,x_2)$.
\Cref{fig:slice} shows the allocation of $M_\eps$ as bidder~1's
translated type varies, for fixed $x_2$ and $\eps\le d$. When the
opponent's total exceeds $\eps$, the allocation is the same as under
$M_0$; when it is below $\eps$, bidder~1 either receives both items
or neither item is sold.

\subsection{Main results}\label{sec:results}

The no-reserve auction has an optimality property independent of the size of the shift. It isolates the allocation problem from the decision whether to sell.

\begin{theorem}[Complete sale]\label{thm:complete-sale}
For every $c\ge0$, $M_0$ maximizes expected revenue among all must-sell DSIC, ex-post IR mechanisms. The optimal revenue is
\[
2c+\frac{637}{810}.
\]
\end{theorem}
Optimality is proved in \cref{sec:complete-sale-proof}, and the
revenue formula is derived in \cref{prf:revenue-no-reserve}.
Under complete sale the shift contributes the constant $2c$ to
revenue. At $c=0$, the revenue agrees with $R(1/3)$ in
\textcite[Example~3]{JMM2007}, since rewarding bundle outcomes by
$d$ is equivalent to penalizing split outcomes by $d$.
The following theorem removes the complete-sale restriction for
sufficiently large shifts.

Let $\cst=91/15$.

\begin{theorem}[Unrestricted sale]\label{thm:main}
For every $c\ge91/15$, $M_\eps$ maximizes expected revenue among all DSIC, ex-post IR mechanisms. The optimal revenue is
\[
2c+\frac{637}{810}+\frac{\eps^3}{12}+\frac{3\eps^5}{40}.
\]
\end{theorem}

Optimality is proved in \cref{sec:reserve-proof}. The revenue
formula, derived in \cref{prf:prop:Mst-revenue}, holds for every
$c\ge5/4$, even when optimality of $M_\eps$ is not established.
In particular, the reserve strictly increases revenue relative to
$M_0$ throughout this range. The threshold $\cst$ is sufficient and
is not claimed to be sharp; it follows from the lower bound on the
virtual valuations established in \cref{sec:threshold}.


\section{Revenue upper bound}\label{sec:field}

\subsection{Candidate utilities and allocation regions}\label{sec:candidate-properties}

The following notation and auxiliary properties are used in the proof.
Write $r^+=\max\{r,0\}$ for $r\in\R$. For $e\in\R^2$ put
\[
\mathcal{U}(e):=\max\bigl(0,\ e_1-d,\ e_2-d,\ e_1+e_2\bigr).
\]
The function $\mathcal U:\R^2\to\R$ gives the no-reserve utility as a function of the bidder's valuation vector minus the opponent's. Define the four allocation regions by
\begin{gather*}
C_{11}=\{e\in\R^2:e_1\ge-d,\ e_2\ge-d,\ \one\cdot e\ge0\},\qquad C_{00}=-C_{11},\\
C_{10}=\{e\in\R^2:e_1\ge d,\ e_2\le-d\},\qquad C_{01}=-C_{10}.
\end{gather*}
These closed convex regions correspond to the maximizing pieces of $\mathcal U$ and cover $\R^2$ with pairwise disjoint interiors and null boundaries. Their subscripts record the bidder's allocation vector. In particular, $\mathcal U=0$ on $C_{00}$, including whenever $e\le0$.

\begin{lemma}[Candidate utilities]\label{lem:candidate-utilities}
For every $c\ge0$ and each bidder $i$, the no-reserve utility is
\[
u_i^0(v)=\mathcal U(v_i-v_{-i}).
\]
For $c\ge5/4$, the reserve utility is
\[
u_i^\eps(v)=
\begin{cases}
\mathcal U(v_i-v_{-i}),&\one\cdot v_{-i}\ge2c+\eps,\\
(\one\cdot v_i-2c-\eps)^+,&\one\cdot v_{-i}<2c+\eps.
\end{cases}
\]
For either mechanism in the stated range, the utility of the
lowest own type $c\one$ is zero for every opponent type $v_{-i}$.
\end{lemma}

The proof is given in \cref{app:proofs-model}.

\paragraph{Allocation regions.}
From this point on, we use the translated coordinates of
\cref{sec:mech} and retain the same symbols for mechanism functions;
for example,
\[
q_i(x):=q_i(c\one+x_1,c\one+x_2).
\]
The same convention applies to payments, utilities, and candidate
functions. Reports are translated as $\hat x_i=\hat v_i-c\one$;
DSIC and IR are unchanged. In particular, own type $0$ corresponds
to the original type $c\one$. We write $(a,x_{-i})$ for the profile
obtained by replacing bidder~$i$'s type by $a\in X$.

For a bidder with translated type $a$ facing an opponent with
translated type $x_{-i}$, define
\[
C_{jk}(x_{-i})=X\cap(x_{-i}+C_{jk}),\qquad (j,k)\in\{0,1\}^2.
\]
These cells partition the type square up to null boundaries and
give the bidder's allocation under $M_0$. In particular,
\[
C_{11}(x_{-i})=\{a\in X:a\ge x_{-i}-d\one,\ \one\cdot a\ge \one\cdot x_{-i}\},
\]
\[
C_{00}(x_{-i})=\{a\in X:a\le x_{-i}+d\one,\ \one\cdot a\le \one\cdot x_{-i}\}.
\]
For $r\ge0$, let $T_r=\{a\in X:\one\cdot a\le r\}$. Under $M_\eps$,
when $\eps\le d$, the allocation is unchanged for an opponent with
$\one\cdot x_{-i}>\eps$; when $\one\cdot x_{-i}<\eps$, the bidder receives both items if
$\one\cdot a>\eps$ and neither item otherwise. At the reserve boundary,
the priority rule favors no sale whenever both totals are at most
$\eps$. Thus the no-sale region is $T_\eps\times T_\eps$.
The geometry is illustrated in \cref{fig:slice}.

\subsection{Overview of the proof}\label{sec:overview}

The proof separates the revenue comparison into two parts: pointwise
virtual-welfare maximization and a conditional utility bound. We
assign the same virtual valuations to both bidders whenever the
candidate awards the bundle to one of them. Under complete sale,
this makes the allocation of the bundle immaterial to virtual
welfare, leaving the conditional utility bound as the main difficulty.
We establish this bound by choosing the field so that integration by
parts reduces it to the region in which the candidate allocates
nothing to the bidder. Individual rationality and coordinatewise
monotonicity of utility then suffice.

With a reserve, the virtual valuations must also be nonnegative on
sold items. To achieve this, we modify the common field while
controlling the resulting change in the conditional utility bound.
The modification exploits monotonicity of the difference in
allocation probabilities along segments of constant total valuation.
Thus the reserve proof uses one additional consequence of incentive
compatibility, rather than joint convexity of utility.

\subsection{Incentive constraints used in the proof}\label{sec:menus}

\begin{lemma}\label{lem:slices}
Let $M$ be DSIC. For each bidder $i$ and each fixed
$x_{-i}\in X$, the utility slice $u_i(\cdot,x_{-i}):X\to\R$ is
coordinatewise nondecreasing and $1$-Lipschitz in the $\ell^1$ norm.
At every differentiable interior point,
\[
\nabla_{x_i}u_i(x)=q_i(x),
\]
and hence the equality holds almost everywhere in the own type.
\end{lemma}
\begin{proof}
Fix bidder $i$ and an opponent type $x_{-i}$, writing
$u(a)=u_i(a,x_{-i})$ for the utility slice. If $a,\hat a\in X$ differ only in
coordinate $j$ and $a_j>\hat a_j$, the two IC inequalities give
\[
q_{ij}(\hat a,x_{-i})\le
\frac{u(a)-u(\hat a)}{a_j-\hat a_j}
\le q_{ij}(a,x_{-i}).
\]
Since allocations lie in $[0,1]$, utility is nondecreasing and
$1$-Lipschitz on each coordinate line. Changing coordinates
successively gives the stated Lipschitz bound. Taking one-sided
difference quotients at differentiable points identifies
$\partial_j u(a)$ with $q_{ij}(a,x_{-i})$; Rademacher's theorem gives
differentiability almost everywhere
\parencite[Section~3.1]{EvansGariepy}.
\end{proof}

With the same slice notation, on each coordinate line, absolute continuity and the same difference
quotient argument give the one-dimensional envelope identity
\begin{equation}\label{eq:coordinate-envelope}
u(r,t)=u(0,t)+\int_0^r q_{i1}((w,t),x_{-i})\,\dd w,
\qquad 0\le r,t\le1,
\end{equation}
and its vertical counterpart. Joint convexity of the utility slice
is not required.

\begin{lemma}[Local swap monotonicity]\label{lem:local-swap}
Let $M$ be DSIC. Fix bidder $i$ and the opponent type $x_{-i}\in X$.
Along any segment in $X$ of constant total valuation, the difference
\[
q_{i1}(x_i,x_{-i})-q_{i2}(x_i,x_{-i})
\]
is nondecreasing in $x_{i1}$.
\end{lemma}

\begin{proof}
For any $a,\hat a\in X$ with $\one\cdot a=\one\cdot\hat a$,
adding the two IC inequalities gives
\[
(a-\hat a)\cdot[q_i(a,x_{-i})-q_i(\hat a,x_{-i})]\ge0.
\]
If $a_1>\hat a_1$, then $a-\hat a=(a_1-\hat a_1)(1,-1)$.
Dividing by $a_1-\hat a_1$ proves the claim.
\end{proof}

\subsection{Revenue-gap decomposition}\label{sec:revenue-gap}

For a measurable integrable field $V:X\times X\to\R^2$ and a
Lipschitz utility slice $u:X\to\R$, define the conditional gap
\begin{equation}\label{eq:slicebound}
D_{x_{-i}}(u;V)=\int_X[u(a)+(V(a;x_{-i})-a)\cdot\nabla u(a)]\,\dd a,
\end{equation}
for each $x_{-i}\in X$ for which the integral is defined.
The payment identity and \cref{lem:slices} give
\begin{equation}\label{eq:revenue-dual}
\begin{split}
\mathrm{Rev}(M)
={}&\mathbb E_x\left[\sum_{i=1}^2
(c\one+V(x_i;x_{-i}))\cdot q_i(x)\right]\\
&-\sum_{i=1}^2\mathbb E_{x_{-i}}
[D_{x_{-i}}(u_i(\cdot,x_{-i});V)] .
\end{split}
\end{equation}
All terms are integrable by the allocation bound and integrability
of the payments and field.

The identity separates the revenue comparison into virtual-welfare
maximization and a conditional utility bound. The following criterion
formalizes this observation and applies to both the complete-sale
problem and the problem with withholding.

\begin{proposition}[Dual optimality criterion]\label{prop:dual-criterion}
Fix the class of DSIC, ex-post IR mechanisms, either with or without
the complete-sale restriction. Let $M^*$ be an admissible mechanism,
with allocation $q^*$ and utility $u^*$. Suppose there exists an
integrable measurable field $V:X\times X\to\R^2$ satisfying:
\begin{enumerate}[label=(\roman*)]
\item For almost every type profile $x$, the candidate allocation
$q^*(x)$ maximizes
\[
\sum_{i=1}^2(c\one+V(x_i;x_{-i}))\cdot q_i
\]
over feasible allocation vectors $(q_1,q_2)$, imposing complete sale
when required.
\item For every admissible mechanism $M$ with utility $u_i$, and for
each bidder $i$,
\[
D_{x_{-i}}(u_i(\cdot,x_{-i});V)\ge0
\]
for almost every $x_{-i}$.
\item For each bidder $i$,
\[
D_{x_{-i}}(u_i^*(\cdot,x_{-i});V)=0
\]
for almost every $x_{-i}$.
\end{enumerate}
Then $M^*$ maximizes expected revenue in the specified class.
\end{proposition}

\begin{proof}
By condition (iii), subtracting \eqref{eq:revenue-dual} for any
admissible mechanism $M$ from the same identity for $M^*$ yields
\begin{equation}\label{eq:revenue-gap}
\begin{split}
\mathrm{Rev}(M^*)-\mathrm{Rev}(M)
={}&\mathbb E_x\left[\sum_{i=1}^2
(c\one+V(x_i;x_{-i}))\cdot(q_i^*(x)-q_i(x))\right]\\
&+\sum_{i=1}^2\mathbb E_{x_{-i}}
[D_{x_{-i}}(u_i(\cdot,x_{-i});V)] .
\end{split}
\end{equation}
The first term, the allocation gap, is nonnegative by condition (i).
The second, the sum of conditional gaps, is nonnegative by condition
(ii).
\end{proof}

We now construct fields $V_0$ and $V_\eps$ satisfying
\cref{prop:dual-criterion} for $M_0$ under complete sale and for
$M_\eps$ when withholding is allowed, respectively.

\subsection{The dual fields and the allocation gap}\label{sec:field-def}

A bundle pair $(h,l)\in X\times X$ consists of a candidate bundle
winner $h$ and loser $l$, satisfying $\one\cdot h\ge \one\cdot l$ and $h_j\ge l_j-d$
for $j=1,2$. For the reserve mechanism, such a pair is sold if
$\one\cdot h>\eps$. All allocation regions are those of the candidate.

We seek common fields $S_0$ and $S_\eps$ on the
bundle pairs of $M_0$ and the sold bundle pairs of $M_\eps$,
respectively. The key feature is that both bidders are assigned
the same field value whenever the candidate sells the bundle to
one bidder.
On split pairs, a bidder with own type $a$ receiving only item $j$
under the candidate is assigned the corresponding field
\begin{equation}\label{eq:item-fields}
V^{(1)}(a)=\tfrac32a-\tfrac12(1,0),\qquad
V^{(2)}(a)=\tfrac32a-\tfrac12(0,1).
\end{equation}
Thus $V_r(a;x_{-i})$, for $r\in\{0,\eps\}$, equals $S_r(h,l)$ on
bundle pairs sold by $M_r$, the corresponding item field on split
pairs, and $-c\one$ on no-sale pairs of $M_\eps$.
Values on null interfaces are immaterial.

For this choice of fields, condition (i) of \cref{prop:dual-criterion}
holds under complete sale. In the reserve problem, it holds provided
$S_{\eps,j}(h,l)\ge-c$ for both items on almost every sold bundle pair.
Indeed, on a sold bundle pair both bidders have the same virtual
valuations. With complete sale their total allocation is fixed;
with withholding, selling both items is optimal if these valuations
are nonnegative. On a no-sale pair both bidders' virtual valuations
are zero. On a split pair with $x_{11}-x_{21}>d$ and
$x_{22}-x_{12}>d$, the differences are
\[
V^{(1)}(x_1)-V^{(2)}(x_2)
=\tfrac32(x_{11}-x_{21}-d,\ x_{12}-x_{22}+d).
\]
They favor the specified recipients, whose virtual valuations for
the respective items are nonnegative since their own translated
valuations exceed $d$. The other split is symmetric.

It remains to establish the conditional gap conditions (ii) and (iii)
of \cref{prop:dual-criterion} and, in the reserve problem, the
stated lower bound on the common field. The construction uses
$S_0$ on the winner side because its divergence and normal traces
allow the winner-region contributions to the conditional gap to
cancel. In the reserve problem we modify this field to ensure
nonnegative virtual valuations. The modification need not preserve
the exact winner-side divergence, so we first bound the conditional
gap using an auxiliary field that retains the unmodified field on
the winner side and uses the modified field on the loser side.

For $r\in\{0,\eps\}$, let $W_r$ equal $S_0(a,x_{-i})$ on own
bundle-winning cells sold by $M_r$ and $V_r(a;x_{-i})$ elsewhere;
in particular, $W_0=V_0$. To pass from $W_\eps$ to $V_\eps$, we
choose the winner-side correction so that local swap monotonicity
makes its contribution to the conditional gap nonnegative, while
it vanishes at the candidate. The following lemma states the
properties required of the construction.

\begin{lemma}[Existence of dual fields]\label{lem:dual-fields}
For every $c\ge0$ there exists a measurable common field $S_0$
on bundle pairs of $M_0$. For $c\ge\cst$, the field can be chosen
jointly with a measurable common field $S_\eps$ on sold bundle
pairs of $M_\eps$. The induced fields $V_r$ and $W_r$ are integrable
and have the following properties in their respective ranges:
\begin{enumerate}[label=(\roman*)]
\item For almost every opponent type $x_{-i}$,
\[
D_{x_{-i}}(u;W_r)\ge0
\]
for every nonnegative coordinatewise nondecreasing Lipschitz
utility slice $u:X\to\R$, with equality at the corresponding
candidate's utility slice.
\item For $c\ge\cst$, for almost every opponent type and almost
every fixed-total level $a=(p,s-p)$, the correction $V_\eps-W_\eps$
is a finite sum of terms of the form $g(p)(1,-1)$, supported on
bundle-winning subsegments with $s\le1$. On each such subsegment,
$g$ is integrable, has zero integral, and is nonpositive below a
cutoff and nonnegative above it.
\item For $c\ge\cst$, $S_{\eps,j}(h,l)\ge-c$ for $j=1,2$ on
almost every sold bundle pair.
\end{enumerate}
\end{lemma}

The construction and proof of \cref{lem:dual-fields} are given in
\cref{sec:reduction,sec:cert,sec:loser,sec:upper-verification}.
Property (i) is obtained by reducing the conditional gap to
the loser or no-sale region; property (ii) permits the use of
\cref{lem:local-swap}, and property (iii) ensures virtual-welfare
maximization when withholding is allowed.


\section{Proof of optimality}\label{sec:corner}

\subsection{Optimality under complete sale}\label{sec:complete-sale-proof}

Consider a must-sell DSIC, ex-post IR mechanism.
By ex-post IR and \cref{lem:slices}, each utility slice is
nonnegative, coordinatewise nondecreasing and Lipschitz.
Property (i) of \cref{lem:dual-fields}, with $W_0=V_0$, gives
\[
D_{x_{-i}}(u_i(\cdot,x_{-i});V_0)\ge0
\]
for almost every $x_{-i}\in X$ and each $i$, with equality for $M_0$.
Thus conditions (ii) and (iii) of \cref{prop:dual-criterion} hold.
Condition (i) follows from the complete-sale comparison in
\cref{sec:field-def}. Applying \cref{prop:dual-criterion} with $M^*=M_0$
and $V=V_0$ proves \cref{thm:complete-sale}.

\subsection{Optimality with a reserve}\label{sec:reserve-proof}

Let $c\ge\cst$ and consider a DSIC, ex-post IR mechanism.
Property (iii) of \cref{lem:dual-fields} and the allocation
comparison in \cref{sec:field-def} establish condition (i) of
\cref{prop:dual-criterion}. By ex-post IR, \cref{lem:slices}, and
property (i) of \cref{lem:dual-fields}, the conditional gap for
$W_\eps$ is nonnegative and is zero at the candidate. It remains
to transfer these properties to $V_\eps$.
Fix bidder $i$ and an opponent type $x_{-i}$ outside the exceptional
null set. Write $u=u_i(\cdot,x_{-i})$ for the utility slice.

By \cref{lem:local-swap}, the difference of allocation
probabilities along each fixed-total segment is nondecreasing
in $p$. If an integrable coefficient $g$ has zero integral on a subsegment
and is nonpositive below a cutoff and nonnegative above it, then
\[
\int g(p)[q_{i1}((p,s-p),x_{-i})-q_{i2}((p,s-p),x_{-i})]\,\dd p\ge0,
\]
where the integral is over that subsegment. To see this, subtract
the value of the nondecreasing factor at the cutoff.
By property (ii) of \cref{lem:dual-fields}, this inequality
applies to each term of the correction $V_\eps-W_\eps$.
Since the change of variables $a=(p,s-p)$ has unit absolute
Jacobian, integrating over the levels and using \cref{lem:slices}
gives
\[
D_{x_{-i}}(u;V_\eps)-D_{x_{-i}}(u;W_\eps)\ge0.
\]
The difference vanishes at the candidate, whose allocation
coordinates coincide on every bundle-winning cell.
Together with the auxiliary gap, this establishes conditions (ii)
and (iii) of \cref{prop:dual-criterion}. Applying that proposition
with $M^*=M_\eps$ and $V=V_\eps$ proves \cref{thm:main}.

\paragraph{Incentive constraints used in the proof.}
The complete-sale proof uses only IC inequalities between own
reports differing in one coordinate. The reserve proof additionally
uses monotonicity of the difference between bidder~$i$'s two
allocation probabilities on sold bundle-winning segments of
translated total $s\le1$.


\section{Concluding remarks}\label{sec:remarks}

We have shown that a deterministic mixed-bundling auction is optimal
under complete sale for every shift $c\ge0$, and that adding a bundle
reserve yields an optimal auction without the complete-sale
restriction for sufficiently large shifts.

A natural next question is how far the latter result extends to
smaller shifts. For $5/4\le c<\cst$, the same mechanism remains
a candidate for optimality. For $c<5/4$, the reserve premium exceeds
the split penalty, so the reserve interacts with split allocations
and the analysis must account for a different menu structure.
In particular, the unshifted benchmark $U[0,1]$ remains unresolved:
the computational upper bound of $0.892$
\parencite{JiangParkesWang2026} exceeds the reported revenues of
approximately $0.876$ from GemNet \parencite{WangJiangParkes2024}
and $0.871$ from the mixed-bundling family
\parencite{TangSandholm2012}.

Two further questions concern the scope of the result. Does the
optimality of mixed-bundling affine maximizers extend to more bidders,
more items, or other distributions? And does Bayesian incentive
compatibility permit strictly higher revenue than dominant-strategy
incentive compatibility in the large-shift environment, as can occur
in the continuous example of \textcite{Yao2017}? The continuous
BIC duality framework of \textcite{KSTZ2022} provides a starting
point for the latter question.


\appendix
\crefalias{section}{appendix}

\section{Proofs for the candidate mechanism}\label{app:proofs}

We first verify the candidate's incentive properties and derive its allocations and utilities, then compute revenue with and without a reserve. This
order separates the incentive properties of the candidate from its
optimality, which is established in the main text using the field verified in the subsequent appendices. We use
the translated coordinates and the mechanism-function convention of
\cref{sec:candidate-properties} throughout.

\subsection{Allocation, payments, and utility}\label{app:proofs-model}\label{prf:prop:Mst}

For completeness, we record the affine-maximizer payment formula and its incentive properties.
An \emph{affine maximizer} is given by a finite set $O$ of \emph{outcomes}, each outcome $o$ being a deterministic feasible allocation $(\bar q_1(o),\bar q_2(o))\in\{0,1\}^2\times\{0,1\}^2$, and an outcome-specific constant $\kappa(o)$.
At reports $\hat v$, the mechanism selects
\[
o^*\in\arg\max_{o\in O}\left[
\sum_{i=1}^2 \hat v_i\cdot\bar q_i(o)+\kappa(o)
\right],
\]
with ties broken by a fixed priority order on $O$. It charges the Clarke-type payments \parencite{Clarke1971,Groves1973,Vickrey1961}
\[
\begin{split}
p_i(\hat v)={}&
\max_{o\in O}\bigl[\hat v_{-i}\cdot\bar q_{-i}(o)+\kappa(o)\bigr]\\
&-\bigl[\hat v_{-i}\cdot\bar q_{-i}(o^*)+\kappa(o^*)\bigr],
\end{split}
\]
where the subscript $-i$ denotes the other bidder. When the constants are all equal, this is the VCG mechanism restricted to the outcomes in $O$; for affine maximizers with outcome-specific constants in multi-object auctions see \parencite{JMM2007} and \parencite[Section~4.2]{CurrySandholmDickerson2023}.

\begin{lemma}\label{lem:AMA}
Every affine maximizer is DSIC and ex-post IR in this nonnegative-valuation environment. Its truthful utility is
\[
\begin{split}
u_i(v)={}&\max_{o\in O}\left[
\sum_{k=1}^2 v_k\cdot\bar q_k(o)+\kappa(o)\right]\\
&-\max_{o\in O}\left[v_{-i}\cdot\bar q_{-i}(o)+\kappa(o)\right],
\end{split}
\]
independent of the tie-breaking rule.
\end{lemma}
\begin{proof}
Substituting the payment formula into bidder~$i$'s utility yields the displayed difference of maxima, since $o^*$ maximizes the first objective. A misreport may change the selected outcome, but the second maximum depends only on the other bidder's report. Evaluated at bidder~$i$'s true valuation, the first objective is maximized by truthful reporting, so a misreport cannot increase utility. This proves DSIC.

For IR, note that $v_i\cdot\bar q_i(o)\ge0$ for every outcome because valuations and allocation probabilities are nonnegative. The first maximum is therefore at least the second.
\end{proof}

For $M_\eps$, the constants are $\kappa(\mathrm B_1)=\kappa(\mathrm B_2)=0$, $\kappa(\mathrm S_{12})=\kappa(\mathrm S_{21})=-d$, and $\kappa(\mathrm N)=2c+\eps$. For $M_0$, the same sale-outcome constants apply, and $\mathrm N$ is omitted.

\paragraph{The reserve premium.}
To derive the reserve, consider a single buyer offered the bundle at price $2c+r$, with $0\le r\le1$. The expected revenue is $(2c+r)(1-r^2/2)$, whose derivative is $1-3r^2/2-2cr$. This derivative vanishes at $r=\eps(c)$, is positive below it, and is negative above it. Prices below $2c$ cannot improve upon selling at $2c$; for $1\le r\le2$, expected revenue is $(2c+r)(2-r)^2/2$, which is decreasing in $r$. Thus $2c+\eps(c)$ is the optimal pure-bundling price for a single buyer with independent $U[c,c+1]$ valuations \parencite[Example~3]{Pavlov2011} \parencite[see also][Theorem~5]{DDT2017}. Solving \eqref{eq:eps} for $c$ gives $c=1/(2\eps)-3\eps/4$, so $c\ge5/4$ if and only if $\eps(c)\le1/3$. The same reserve equation determines the virtual-value floor in the proof of optimality (\cref{sec:corner}).

\begin{proof}[Proof of \cref{lem:candidate-utilities} and the allocation properties]
We use the original valuations $v_i,v_{-i}$ in $X_c$.

\emph{Without a reserve.}
In $M_0$, the opponent's adjusted welfare is maximized by assigning both items to the opponent, giving $\one\cdot v_{-i}$. The payment formula therefore gives prices $0$, $v_{-i,j}+d$ and $\one\cdot v_{-i}$ for nothing, item $j$ and both items, respectively. Maximizing utility over this menu yields
\[
u_i^0(v)=\mathcal U(v_i-v_{-i}).
\]
The allocation follows from the maximizing pieces of $\mathcal U$ and the definitions of the four regions. Every outcome sells both items, so $M_0$ is must-sell.

\emph{With a reserve.}
Adding the no-sale outcome introduces the same additional value $2c+\eps$ into both maxima in \cref{lem:AMA}. Consequently,
\[
\begin{aligned}
u_i^\eps(v)
&=\max\{\one\cdot v_{-i}+u_i^0(v),2c+\eps\}
  -\max\{\one\cdot v_{-i},2c+\eps\}\\
&=\bigl[u_i^0(v)-(2c+\eps-\one\cdot v_{-i})^+\bigr]^+.
\end{aligned}
\]
By the payment formula, the price of every nonempty bundle increases by $(2c+\eps-\one\cdot v_{-i})^+$. Receiving nothing still costs zero: among the outcomes assigning nothing to the bidder, the one maximizing the opponent's adjusted welfare attains $\max\{\one\cdot v_{-i},2c+\eps\}$. This gives the stated menu.

If $\one\cdot v_{-i}\ge2c+\eps$, the utility is unchanged. If $\one\cdot v_{-i}\le2c+\eps$, then $v_{-i,j}\le c+\eps\le c+d$ for both items, while $v_{ij}\ge c$. Hence each single-item utility in the no-reserve menu is at most the utility from buying both items. It follows that $u_i^0(v)=(\one\cdot v_i-\one\cdot v_{-i})^+$, and thus $u_i^\eps(v)=(\one\cdot v_i-2c-\eps)^+$ in this case.

The no-sale outcome can maximize adjusted welfare only if both bidders' totals are at most $2c+\eps$. Conversely, if both totals satisfy this bound, the preceding comparison gives $\one\cdot v_{-i}+u_i^0(v)=\max\{\one\cdot v_i,\one\cdot v_{-i}\}\le2c+\eps$. The no-sale outcome is then selected by the fixed priority rule. Outside this region, the maximizer is selected from the original four outcomes using the same priority order as in $M_0$.

Both mechanisms are DSIC and ex-post IR by \cref{lem:AMA}. At $v_i=c\one$, we have $\mathcal U(v_i-v_{-i})=0$, so the displayed relation also gives $u_i^\eps(v)=0$.
\end{proof}

\subsection{The revenue identity}\label{sec:revid}

The payment is $p_i(x)=q_i(x)\cdot(c\one+x_i)-u_i(x)$, and by \cref{lem:slices} $q_i(x)=\nabla_{x_i}u_i(x)$ almost everywhere in the own type.
Hence the payment is a function of the utility alone (compare the envelope theorem of \parencite{MilgromSegal2002}), and an integration by parts removes the gradient.
The result is expressed through the following signed measure on $X$:
\[
\mu:=\delta_0-3\,\Leb\lfloor X+\Hone\lfloor\mathrm{top}+\Hone\lfloor\mathrm{right},
\qquad \mathrm{top}:=[0,1]\times\{1\},\quad\mathrm{right}:=\{1\}\times[0,1],
\]
where $\delta_0$ is the unit point mass at the origin, $\Leb$ is the Lebesgue measure on $\R^2$ and $\Hone$ is the length measure on segments.
Its total mass is $\mu(X)=1-3+1+1=0$.
The measure $\mu$ plays the role of the ``transformed measure'' of \textcite[Section~2.2]{DDT2017} for the uniform density.

\begin{lemma}\label{lem:revenue}\leavevmode
\begin{enumerate}[label=(\alph*)]
\item For every Lipschitz function $f$ on $X$,
\[
\int_X\bigl(a\cdot\nabla f(a)-f(a)\bigr)\dd a=\int f\dd\mu-f(0).
\]
\item For every DSIC mechanism $M$,
\begin{equation}\label{eq:rev}
\begin{split}
\mathrm{Rev}(M)={}&c\,\mathbb E_x\left[\sum_{i=1}^2 q_i(x)\cdot\one\right]\\
&+\sum_{i=1}^2\int_X\left[\int_X u_i(a,x_{-i})\dd\mu(a)-u_i(0,x_{-i})\right]\dd x_{-i}.
\end{split}
\end{equation}
\end{enumerate}
\end{lemma}

\begin{proof}
For part (a), fix $a_2\in[0,1]$. Since $f$ is Lipschitz, $f(\cdot,a_2)$ is absolutely continuous, and integration by parts gives
\[
\int_0^1 a_1\partial_1 f(a_1,a_2)\dd a_1
=f(1,a_2)-\int_0^1 f(a_1,a_2)\dd a_1.
\]
By Rademacher's theorem \parencite[Section~3.1]{EvansGariepy} the gradient exists almost everywhere. Integrate over $a_2$ and repeat with the coordinates exchanged. Fubini's theorem yields
\[
\int_X a\cdot\nabla f\dd a=
\int_{\mathrm{right}}f\dd\Hone+\int_{\mathrm{top}}f\dd\Hone-2\int_X f\dd a.
\]
Subtract $\int_X f\dd a$ and use the definition of $\mu$ to obtain part (a).

For part (b), fix $i$ and $x_{-i}$. By \cref{lem:slices}, the payment identity becomes
\[
p_i(x)=c\,q_i(x)\cdot\one+x_i\cdot\nabla_{x_i}u_i(x)-u_i(x)
\]
almost everywhere in $x_i$. Applying part (a) to the Lipschitz slice $u_i(\cdot,x_{-i})$ gives
\[
\begin{split}
\int_Xp_i(x)\dd x_i
={}&c\int_X q_i(x)\cdot\one\dd x_i\\
&+\int_X u_i(a,x_{-i})\dd\mu(a)-u_i(0,x_{-i}).
\end{split}
\]
Both sides are integrable in $x_{-i}$ by the payment and allocation bounds and Fubini. Integrate over $x_{-i}$ and sum over $i$ to obtain \eqref{eq:rev}.
\end{proof}

For a must-sell mechanism, the first term of \eqref{eq:rev} is the constant $2c$. The remaining terms are linear in the utility slices. The subtraction of the utility at the lowest type is valid regardless of whether that utility is zero; ex-post IR only makes it nonnegative.

\subsection{Revenue without a reserve}\label{prf:revenue-no-reserve}

\begin{proof}[Derivation of the revenue formula in \cref{thm:complete-sale}]
Let $e_j=x_{1j}-x_{2j}$ for $j=1,2$, and let $U=1-x_{22}$.
The variables $e_1,e_2$ are independent with density $1-|r|$ on
$[-1,1]$, while $U$ is uniform on $[0,1]$ and independent of $e_1$.
By \eqref{eq:rev}, the zero utility at the lowest type, and symmetry
across bidders and items,
\[
\mathrm{Rev}(M_0)-2c=2(-3I_A+2I_T),
\]
where $I_A=\mathbb E\,\mathcal U(e_1,e_2)$ is the area expectation and
$I_T=\mathbb E\,\mathcal U(e_1,U)$ is the top-edge expectation.
The right edge contributes the same expectation as the top edge.

Decompose the utility into the pure-bundling part and the gains from
split allocations:
\[
\mathcal U(e)=(e_1+e_2)^++m_1(e)+m_2(e),
\]
where
\[
m_1(e)=\min\{(e_1-d)^+,(-e_2-d)^+\},\qquad
m_2(e)=\min\{(e_2-d)^+,(-e_1-d)^+\}.
\]
Indeed, both gains vanish on $C_{11}$ and $C_{00}$; on $C_{10}$,
$m_2=0$ and $m_1=e_1-d-(e_1+e_2)^+$, with the symmetric identity
on $C_{01}$.

For the bundle part, let $\mathrm{sl}(t)=\mathbb E(e_1-t)^+$.
Integrating the triangular density gives
$\mathrm{sl}(t)=(1-t)^3/6$ for $0\le t\le1$ and, by symmetry,
$\mathrm{sl}(-t)=t+\mathrm{sl}(t)$. Conditioning on the second
argument yields
\begin{align*}
\mathbb E(e_1+e_2)^+
&=\int_0^1\left[r+\frac{(1-r)^3}{3}\right](1-r)\dd r
=\frac7{30},\\
\mathbb E(e_1+U)^+
&=\int_0^1\left[r+\frac{(1-r)^3}{6}\right]\dd r
=\frac{13}{24}.
\end{align*}

For the split gains, use
$\mathbb E\min(A,B)=\int_0^\infty\Pr(A>t)\Pr(B>t)\dd t$
for independent nonnegative random variables $A,B$.
For $0\le t\le2/3$, the tail probabilities of $e_1-d$ and $U-d$
are $(2/3-t)^2/2$ and $2/3-t$, respectively, and vanish above $2/3$.
Consequently,
\begin{align*}
\mathbb E\,m_1(e_1,e_2)
=\mathbb E\,m_2(e_1,e_2)
&=\int_0^{2/3}\frac{(2/3-t)^4}{4}\dd t
=\frac8{1215},\\
\mathbb E\,m_2(e_1,U)
&=\int_0^{2/3}\frac{(2/3-t)^3}{2}\dd t
=\frac2{81}.
\end{align*}
Since $m_1(e_1,U)=0$, the contributions are summarized by
\begin{center}
\begin{tabular}{lcc}
\toprule
 & $I_A$ & $I_T$\\
\midrule
Pure bundling & $7/30$ & $13/24$\\
Split gains & $16/1215$ & $2/81$\\
\bottomrule
\end{tabular}
\end{center}
Substitution gives $\mathrm{Rev}(M_0)=2c+637/810$.
\end{proof}

\subsection{Revenue with a reserve}\label{prf:prop:Mst-revenue}

\begin{proof}[Derivation of the revenue formula in \cref{thm:main}]
Let $c\ge5/4$. Relative to $M_0$, the reserve withholds both items on
$T_\eps\times T_\eps$, a set of probability $\eps^4/4$. The
shift term in \eqref{eq:rev} therefore falls by $c\eps^4/2$.
The utilities at the lowest type remain zero.

By \cref{lem:candidate-utilities}, the utility slice of bidder~$i$
changes only when $\one\cdot x_{-i}<\eps$. In this case $x_{-i,j}<d$ for
both items, so the bundle term in $\mathcal U(x_i-x_{-i})$
dominates both single-item terms. The utility thus changes from
$(\one\cdot x_i-\one\cdot x_{-i})^+$ to $(\one\cdot x_i-\eps)^+$.
For $0\le r\le1$, define
\[
g(r)=\int_X(\one\cdot a-r)^+\dd\mu(a).
\]
Using the density $t$ of $\one\cdot a$ on $[0,1]$ under the uniform law on
$X$, the area integral is
\[
\int_X(\one\cdot a-r)^+\dd a
=1-r+\int_0^r(r-t)t\dd t
=1-r+\frac{r^3}{6}.
\]
Each upper-edge integral equals $3/2-r$, and the atom at the
origin contributes zero. Hence
\[
g(r)=-3\left(1-r+\frac{r^3}{6}\right)
+2\left(\frac32-r\right)=r-\frac{r^3}{2}.
\]
Since $\one\cdot x_{-i}$ has density $t$ on $[0,\eps]$, the change in
the utility integral for each bidder is
\[
\int_0^\eps[g(\eps)-g(t)]t\dd t
=\frac{\eps^3}{6}-\frac{3\eps^5}{20}.
\]
Combining these terms in \eqref{eq:rev} gives
\[
\mathrm{Rev}(M_\eps)-\mathrm{Rev}(M_0)
=-\frac{c\eps^4}{2}+\frac{\eps^3}{3}
-\frac{3\eps^5}{10}.
\]
Substituting $4c\eps=2-3\eps^2$ from \eqref{eq:eps} yields
\[
\mathrm{Rev}(M_\eps)=2c+\frac{637}{810}
+\frac{\eps^3}{12}+\frac{3\eps^5}{40}.\qedhere
\]
\end{proof}

\section{Conditional gap on the loser region}\label{sec:reduction}

This appendix describes the candidate allocation cells and derives
the conditional gap using the boundary and divergence data of the
outer field. The sign check reduces to row and column conditions on
the auxiliary field, not additional assumptions on the model.

\subsection{Allocation boundaries and split fields}\label{sec:cells}

Use the allocation cells $C_{jk}(x_{-i})$ defined in \cref{sec:candidate-properties}.
For a bundle pair $(h,l)$ as defined in \cref{sec:field-def},
the possible winners facing $l$ form $C_{11}(l)$, and the possible
losers facing $h$ form $C_{00}(h)$. The latter is a closed convex
down-set containing $0$, and the no-reserve utility vanishes on it.
We call $C_{00}(h)$ the loser cell of $h$.

\paragraph{Ties and faces.}
For the integration by parts below, the relevant interfaces are the following segments. The bundle tie is
\[
C_{11}(l)\cap C_{00}(l)=\{h\in X:\ \one\cdot h=\one\cdot l,\ |h_1-l_1|\le d\},
\]
the set of types that tie with $l$ for the bundle. The \emph{faces} of $C_{11}(l)$ are
\begin{align*}
C_{11}(l)\cap C_{01}(l)&=\{h\in X:\ h_1=l_1-d,\ h_2\ge l_2+d\},\\
C_{11}(l)\cap C_{10}(l)&=\{h\in X:\ h_2=l_2-d,\ h_1\ge l_1+d\},
\end{align*}
and the \emph{loser faces} of $C_{00}(x_{-i})$ are
\begin{align*}
C_{00}(x_{-i})\cap C_{10}(x_{-i})&=\{a\in X:\ a_1=x_{-i,1}+d,\ a_2\le x_{-i,2}-d\},\\
C_{00}(x_{-i})\cap C_{01}(x_{-i})&=\{a\in X:\ a_2=x_{-i,2}+d,\ a_1\le x_{-i,1}-d\}.
\end{align*}
These equalities follow directly from the definitions of the difference regions; also $C_{10}(x_{-i})\cap C_{01}(x_{-i})=\emptyset$.
A loser face is a face seen from the other side: $a\in C_{00}(x_{-i})\cap C_{10}(x_{-i})$ iff $x_{-i}\in C_{11}(a)\cap C_{01}(a)$, since both say $x_{-i,1}=a_1-d$ and $x_{-i,2}\ge a_2+d$.
The item cells meet the boundary of $X$ in only two edges each: $a\in C_{10}(x_{-i})$ implies $a_1\ge x_{-i,1}+d>0$ and $a_2\le x_{-i,2}-d<1$, so $C_{10}(x_{-i})$ meets $\partial X$ only in the right and bottom edges; by the mirror argument $C_{01}(x_{-i})$ meets it only in the left and top edges.

\paragraph{The item fields.}
The item fields in \eqref{eq:item-fields} have the following divergence and boundary values:
\begin{enumerate}[label=(V\arabic*)]
\item $\dv V^{(1)}=\dv V^{(2)}=\frac32+\frac32=3$.
\item On the edges of $X$ met by the cells: $V^{(1)}_1(1,a_2)=\frac32-\frac12=1$ on the right edge and $V^{(1)}_2(a_1,0)=0$ on the bottom edge; $V^{(2)}_1(0,a_2)=0$ on the left edge and $V^{(2)}_2(a_1,1)=\frac32-\frac12=1$ on the top edge.
\item On the faces: $V^{(1)}_2=\frac32a_2$ on $C_{11}(x_{-i})\cap C_{10}(x_{-i})$ and $V^{(2)}_1=\frac32a_1$ on $C_{11}(x_{-i})\cap C_{01}(x_{-i})$ (true everywhere, by definition).
\item On the loser faces: on $C_{00}(x_{-i})\cap C_{10}(x_{-i})$, $a_1=x_{-i,1}+d$, so $V^{(1)}_1=\frac32x_{-i,1}+\frac12-\frac12=\frac32x_{-i,1}$; on $C_{00}(x_{-i})\cap C_{01}(x_{-i})$, $V^{(2)}_2=\frac32x_{-i,2}$ likewise.
\end{enumerate}
(V1) and (V2) say that on each item cell the field has the divergence and the boundary values that match the measure $\mu$; (V3) and (V4) are the values that the field on the bundle cells is required to take on the faces and the loser faces (see \cref{sec:propH}), so that the fluxes across them cancel.

\subsection{Row and column gap decomposition}\label{sec:propH}
For the auxiliary field $W_r$ of \cref{sec:field-def}, it suffices
that the exact winner field
have divergence $3$ and matching normal traces, with boundary
components $0$ at $h_j=0$, $1$ at $h_j=1$ and $3h_j/2$ at
$h_j=l_j-d$. On ties its component sum must be reciprocal in
$h,l$. The actual field $S_r$ must have the same tie sum and
split-face component.

On a smooth polygonal piece $A$ with outward unit normal $n$,
integration by parts gives
\begin{equation}\label{eq:polygon-residual}
\int_A[u+(V-a)\cdot\nabla u]\,\dd a
=\int_A(3-\dv V)u\,\dd a
+\int_{\partial A}u(V-a)\cdot n\,\dd\Hone.
\end{equation}
Shared interface terms cancel when normal traces agree.
For Lipschitz $u$ the formula follows by uniform approximation
with uniformly bounded gradients converging almost everywhere.
For piecewise smooth fields with jumps, the interior divergence
is distributional, equivalently obtained by summing the polygonal
identities including the normal jump terms.

At the winner and loser split faces the matching components are
$3h_j/2$ and $3l_j/2-1/2=3h_j/2$, respectively. At equal-total
ties the normal is parallel to $\one$, so reciprocity cancels the
terms. At the reserve face, the exact winner component sum is $-2c$,
matching the no-sale field. The winner and split
external boundary terms vanish, as do their volume contributions.
For $r=0$, or for $r=\eps$ and $\one\cdot x_{-i}>\eps$, put $h=x_{-i}$. The gap is
\begin{equation}\label{eq:loser-residual}
\begin{split}
D_{x_{-i}}(u;W_r)={}&\int_{C_{00}(h)}u\,\dd(3\Leb-\dv_l S_r)\\
&+\sum_{j=1}^2\left[
-\int_{C_{00}(h)\cap\{l_j=0\}}S_{r,j}u\,\dd\Hone
+\int_{C_{00}(h)\cap\{l_j=1\}}(S_{r,j}-1)u\,\dd\Hone
\right].
\end{split}
\end{equation}
The divergence measure is on the cell interior and boundary
values are one-sided traces. If $r=\eps$ and $\one\cdot x_{-i}<\eps$, the
analogous residual is on $T_\eps$ with constant field $-c\one$,
hence is nonnegative by IR. No normalization of $u$ is needed.
Exceptional opponent levels may be omitted.

For loser triangles with $\one\cdot h<1/6$, the signs $S_{r,j}\le0$
and $3-\dv_l S_r\ge0$ prove the gap nonnegative directly by IR.
Otherwise use the following criterion.

\begin{lemma}[Row and column criterion]\label{prop:H}
Decompose the interior deficit measure into horizontal and
vertical measures disintegrating along rows and columns.
On almost every horizontal row $l_2=t$ with endpoint
$R_t=\min(1,h_1+d,\one\cdot h-t)$, suppose its horizontal measure
$\mu_t$ is nonnegative, and
\[
\mu_t((0,R_t))+\omega_t\ge S_{r,1}(h,(0,t)),\qquad \omega_t\ge0,
\]
where $\omega_t=S_{r,1}(h,(1,t))-1$ if $R_t=1$ and zero
otherwise. Use one-sided endpoint traces. Impose the mirrored
column conditions. Then $D_{x_{-i}}(u;W_r)\ge0$ for every nonnegative
coordinatewise nondecreasing Lipschitz utility, with equality
at the candidate.
\end{lemma}
\begin{proof}
The row contribution is
\[
\int_{(0,R_t)}u(l_1,t)\,\dd\mu_t(l_1)
+\omega_t u(R_t,t)-S_{r,1}(h,(0,t))u(0,t).
\]
Since $u(l_1,t)\ge u(0,t)$, it is at least
$[\mu_t((0,R_t))+\omega_t-S_{r,1}(h,(0,t))]u(0,t)\ge0$.
Integrate over rows and columns. The candidate utility vanishes
throughout the loser cell.
\end{proof}

The pure horizontal measure is $\frac32\,\dd l_1-\dd_{l_1}S_{r,1}$,
where the second term is the Stieltjes derivative on the open row.
A downward jump of $S_{r,1}$ adds a nonnegative atom.
Its mass is $3R_t/2-S_{r,1}(R_t-)+S_{r,1}(0+)$.
With a nonnegative measure, a tie or split endpoint only
requires $3R_t/2-S_{r,1}(R_t-)\ge0$. A split endpoint has slack
$3d/2=1/2$ by its normal data. At a physical right edge it
suffices that $S_{r,1}(1-)\ge1$; adding the right-edge excess
gives slack $1/2$. The column statements are analogous.
In the inner upper cap we instead split the nonnegative total
deficit equally after the row or column has paid its lower-axis
coefficient.


\section{Uniform kernel and middle strip}\label{sec:cert}

We first specify how the reserve field differs from the no-reserve
field, then construct and verify both fields region by region.

\subsection{The reserve correction}\label{sec:reserve-field}

When withholding is allowed, the common field must also satisfy the
virtual-value floor. We adjust the exact field in the direction
$(1,-1)$, preserving its component sum and hence its normal flux
across equal-total ties. The correction will be chosen so that the
fixed-total monotonicity argument of \cref{sec:reserve-proof}
controls the additional conditional gap.

Set $\delta=1/6$, the half-width of the uniform kernel, and define
$\phi(s)=3s/2-1/s$ for $s>0$. On sold bundle pairs of $M_\eps$,
for $c\ge\cst$, set
\[
S_\eps(h,l)=
\begin{cases}
S_0(h,l)+\phi(\one\cdot h)\left(\tfrac12\one-\tfrac{h}{\one\cdot h}\right),
&\eps<\one\cdot h<\delta,\\
S_0(h,l)+\tau(h,l)(1,-1),&\delta\le \one\cdot h\le1,\\
S_0(h,l),&\one\cdot h>1.
\end{cases}
\]
Here $S_0$ is constructed below, and the scalar correction $\tau$
is specified in \cref{sec:middle-correction}. The component sum
is unchanged by either correction.

On own bundle-winning levels $\eps<s<\delta$ with $s>\one\cdot x_{-i}$,
the entire segment $a=(p,s-p)$, $0\le p\le s$, lies in the
bundle-winning cell since $s<d$. The correction is
\[
-\phi(s)(p/s-1/2)(1,-1).
\]
Its scalar coefficient has zero integral and changes sign from
negative to positive at $p=s/2$, since $\phi(s)<0$.
The corresponding property on the middle strip is verified in
\cref{sec:middle-correction}; no correction is needed above total one.

\subsection{The uniform kernel}

We use a clipped linear profile to construct the stream. Its
reflection identity will ensure reciprocity on bundle ties, and
its saturation before the split threshold $d$ will simplify the
boundary traces. The slope bound will also control the loser-side
row derivatives.

Define $P:\R\to[0,1/2]$ by clipping $1/4-3t/2$ to $[0,1/2]$. Put $k=-P'$ almost everywhere, with $k=3/2$ for $|t|<\delta$ and zero for $|t|\ge \delta$. Thus $P(t)+P(-t)=1/2$, $k$ is even and nonincreasing on the positive half-line.
To adjust the stream near the physical axes without affecting
its interior formula away from the axes, we use a boundary profile
$G$ supported on $[0,\delta]$:
\[
\int_0^\delta P(t)\,\dd t=\frac1{48},\qquad G(t)=(1-6t)_+^2\quad(t\ge0).
\]

For nonnegative $t$, $G(t)=1-48\int_0^tP(w)\,\dd w$ and
$G'(t)=-48P(t)$. In particular $G=0$ above $\delta$. Also

\[
\psi(e)=\frac12-\frac32e-P(e)\ge0\qquad(|e|\le d).
\]

For $|e|<\delta$, $\psi(e)=1/4$; outside that interval the inequality is immediate from saturation.

The winner-side divergence must equal $3$. Rather than impose
this condition directly on both field components, we start from
an affine field with that divergence and add a rotated gradient.
For each fixed loser, a continuous piecewise smooth scalar stream
$\zeta$ thus produces the exact field
\begin{equation}\label{eq:stream}
S_0=\tfrac32h-\tfrac12\one+(-\partial_{h_2}\zeta,\partial_{h_1}\zeta).
\end{equation}
Its winner divergence is $3$ in the distributional sense by
commutation of mixed derivatives. On an interface, the normal component
of the rotated gradient is a tangential derivative of the stream;
matching stream traces therefore match normal fluxes.

\subsection{Middle stream and winner boundary data}

Let $s=\one\cdot h$. For $\delta\le s\le1$, we build the stream from a valuation-difference
profile and two boundary corrections involving $G$. The latter
are localized near the winner axes; taking the difference of the
two item-swapped expressions gives the required item symmetry
of the induced field. On this strip use the stream

\[
\begin{aligned}
\alpha(h_1,l_1;s)={}&(1-s+h_1)P(h_1-l_1)\\
&-P(s-l_1)G(s-h_1)+(1-s)P(l_1)G(h_1),
\end{aligned}
\]

\[
\zeta(h;l)=\alpha(h_1,l_1;s)-\alpha(h_2,l_2;s).
\]

The winner data follow directly: on $h_1=0$, $\zeta=(1-h_2)/2$ by $P(-l_1)+P(l_1)=1/2$ and $G(s)=0$. On $h_1=l_1-d$, the same trace holds because $d>\delta$ and the relevant profile and boundary-correction terms saturate. The mirrored faces follow by item exchange. On a tie, $l_2=s-l_1$ and the stream reduces to

\[
\begin{aligned}
\zeta={}&(2-s)[P(h_1-l_1)+P(l_1)G(h_1)-P(l_2)G(h_2)]\\
&-(1-h_1)/2.
\end{aligned}
\]

Differentiating along the winner level, holding $s$ and $l$ fixed, gives
\[
\frac{\dd}{\dd h_1}\zeta((h_1,s-h_1);l)=\frac12-(2-s)
[k(h_1-l_1)+48P(h_1)P(l_1)+48P(h_2)P(l_2)].
\]
It is reciprocal in $h_1,l_1$ by evenness of $k$. The component
sum of \eqref{eq:stream} is $3s/2-1$ plus this level derivative, so this
proves tie reciprocity.

Direct differentiation gives, almost everywhere,

\[
\begin{aligned}
S_{0,1}={}&3h_1/2-1/2+P(h_1-l_1)-(1-h_1)k(h_2-l_2)\\
&+[P(l_1)+k(s-l_2)]G(h_1)-[P(l_2)+k(s-l_1)]G(h_2)\\
&-48[P(s-l_1)+(1-s)P(l_2)]P(h_2).
\end{aligned}
\]

All second components are obtained by item exchange.

\subsection{Middle loser-side conditions}

At fixed $h,l_2$, the smooth row derivative is

\[
\partial_{l_1} S_{0,1}=k(h_1-l_1)-k(l_1)G(h_1)-48k(s-l_1)P(h_2)\le3/2.
\]

The only jump in $l_1$ comes from $-k(s-l_1)G(h_2)$, and is downward. Hence the pure row deficit is a nonnegative measure. On ties the slack $\Delta=3l_1/2-S_{0,1}$ equals

\[
\begin{aligned}
\Delta={}&\psi(h_1-l_1)+(1-h_1)k(h_1-l_1)\\
&+(P+k)(l_2)G(h_2)-(P+k)(l_1)G(h_1)\\
&+48(2-s)P(l_2)P(h_2).
\end{aligned}
\]

If $h_1\ge \delta$ or $l_1\ge \delta$, the negative term vanishes. Otherwise $k(h_1-l_1)=k(l_1)=3/2$, $\psi(h_1-l_1)=1/4$ and
$P(l_1)\le1/4$. With $t=6h_1\in[0,1]$, the sum of the first,
second and negative terms is at least
\[
\tfrac14+\tfrac32(1-t/6)-\tfrac74(1-t)^2
=\frac{t(13-7t)}4\ge\frac32t.
\]

Split endpoints have slack $3d/2=1/2$ by the prescribed face data. There are no positive-length physical upper edges for these loser cells when $s<1$. The row criterion applies.

\subsection{Angular correction for the reserve}\label{sec:middle-correction}

The reserve correction must have zero component sum and be
nonnegative in its contribution to the winner's conditional gap.
On the middle strip we also need to preserve the loser row
deficit and the tie-endpoint slack. We therefore use a profile
with zero integral and a single sign change, whose positive and
negative parts can be controlled by the bounds in the preceding
subsection.

Define $f:[0,\infty)\to\mathbb R$ by

\[
f(t)=\frac35(1-t)(1-3t)\quad(0\le t\le1),\qquad f(t)=0\quad(t>1).
\]

It has zero integral on $[0,1]$, changes sign once from positive to negative, and satisfies

\[
f(t)\le(1-t)^2,\qquad -f(t)\le t/3\quad(0\le t\le1).
\]

The first inequality follows by factoring $1-t$; the second follows from $27t^2-31t+9>0$ (discriminant $-11$).

We localize this profile near the ends of fixed-total segments,
using the kernel to activate it only for the relevant loser
types. The item-swapped term gives the correction for the other
end. Set
\[
\begin{aligned}
\tau={}&[12P(s-l_1)+k(s-l_1)]f(6h_2)\\
&-[12P(s-l_2)+k(s-l_2)]f(6h_1).
\end{aligned}
\]

In the reserve problem use $S_\eps=S_0+\tau(1,-1)$ on this strip. This correction is valid on winner slices: if its first bracket is nonzero then $l_1>s-\delta$ and
$l_2<\delta$, since $l_1+l_2\le s$. The entire segment
$h_1\in[s-\delta,s]$ lies in the bundle-winning cell:
$h_1\ge s-\delta\ge l_1-d$ and $h_2\ge0\ge l_2-d$. On that segment $f(6h_2)$ has zero integral and changes sign from negative to positive as $h_1$ increases. The mirror term is analogous on $[0,\delta]$. \cref{sec:reserve-proof} therefore applies, regardless of whether the segments overlap. At split faces $\tau=0$: for instance if $h_1=l_1-d$, then $h_2\ge d$ and $s-l_2\ge d$, so both terms vanish. The correction has zero component sum everywhere.

The correction preserves loser row signs. On a smooth row,

\[
\partial_{l_1} S_{\eps,1}=k(h_1-l_1)-k(l_1)G(h_1)-k(s-l_1)[48P(h_2)-12f(6h_2)]\le3/2,
\]

because $f(6h_2)\le G(h_2)\le4P(h_2)$ on the support, and both are zero above $\delta$. The jump of $k(s-l_1)$ now has coefficient $f(6h_2)-G(h_2)\le0$, so it is still downward.

On a tie, the positive part of the first term of $\tau$ is paid for by $(P+k)(l_2)G(h_2)+48(2-s)P(l_2)P(h_2)$ in the exact-field slack. A positive contribution from the mirror term is possible only when $h_1,l_1<\delta$ and $f(6h_1)<0$. It is bounded by $(9/2)(6h_1)/3$, which is at most the lower bound $t(13-7t)/4$ just proved. Thus $3l_1/2-S_{\eps,1}\ge0$. Split endpoints are unchanged. This verifies the entire corrected middle loser region without a polynomial coefficient test.

\subsection{Middle virtual-value floor}\label{sec:middle-floor}

We show that each component is greater than $-91/15$. If $h_2\ge \delta$, the negative terms involving $h_2$ vanish, and the only possibly negative new mirror term is bounded below by $-9/5$. Thus $S_{\eps,1}\ge3h_1-2-9/5$.

For $h_2<\delta$ put $y=1-6h_2\in[0,1]$. Then $G(h_2)=y^2$, $P(h_2)=y/4$, and

\[
f(6h_2)=\frac35y(3y-2).
\]

In the formula for $S_{\eps,1}$ combine $P(h_1-l_1)\ge P(s-l_1)$ with the negative $P(s-l_1)$ term. Use $P(l_2)\le1/4$, $k\le3/2$, and $G(t)-f(6t)\ge0$.
Apart from the possibly negative mirror term $-12P(s-l_2)f(6h_1)$,
the lower bound is
\[
3h_1-2-\tfrac14[G(h_2)+48(1-s)P(h_2)]
-\tfrac32[G(h_2)-f(6h_2)]
-\tfrac14[48P(h_2)-12f(6h_2)-1]_+.
\]
Substituting $y$ gives

\[
3s-\frac52+\left(3s-\frac{43}{10}\right)y+\frac{19}{20}y^2-\frac14\left[\frac{132}{5}y-\frac{108}{5}y^2-1\right]_+.
\]

It is increasing in $s$. Since $s\ge \delta$, when the positive part is active the lower bound is at least

\[
-\frac74-\frac{52}{5}y+\frac{127}{20}y^2\ge-\frac74-\frac{2704}{635}>-\frac{91}{15}.
\]

The first inequality follows by completing the square. When the positive part is inactive the bound is at least $-97/20$.

If $h_1\ge \delta/3$, the retained mirror term is nonnegative. If $h_1<\delta/3$, then $6h_1\ge y$ and $y<1/3$, since $s\ge \delta$. The function $f$ decreases on $[0,1/3]$, so the mirror term is at least $-3f(y)$. In the active-positive-part case the combined bound is then at least

\[
-\frac{71}{20}-\frac{16}{5}y+\frac{19}{20}y^2\ge-\frac{71}{20}-\frac{16}{15}>-\frac{91}{15}.
\]

In the inactive case use the crude bound $-2-(19/5)y-9/5\ge-76/15$. This proves the floor on the entire middle strip.

\section{Lower corner}\label{sec:loser}

Let $s=\one\cdot h$. For $0<s\le\delta$, use the function $\phi$ of
\cref{sec:reserve-field}. In angular coordinates $\theta=h_1/s$, $\eta=l_1/s$, $\nu=l_2/s$ we have $\eta+\nu\le1$.

Near the origin the winner and loser cells have no split faces.
A radial field with component sum $\phi(s)$ has the appropriate
axis traces and a reciprocal tie sum; equalizing its components
in the reserve problem will match the virtual-value floor at
the reserve. We use the associated radial stream as the starting
point and join it to the middle stream.

At $s=\delta$ the discrepancy between the middle stream and the
radial stream $(1-s)(1-2\theta)/2$ factors exactly as

\[
Z=\frac{11}{24}\theta(1-\theta)(\eta-\nu).
\]

Indeed, at $\delta$ the profile and boundary function are linear and quadratic, respectively: $P(\delta(\theta-\eta))=(1-\theta+\eta)/4$ and $G(\delta\theta)=(1-\theta)^2$. Substituting in the middle strip's $\alpha$ and subtracting the radial stream gives the displayed product. The formula follows by expansion and cancellation.

To match the middle stream while leaving the radial formula
unchanged near the reserve, we multiply the discrepancy by a
continuous taper. The quadratic choice below also gives simple
bounds on its derivative. Use
\[
\beta(s)=0\quad(0<s\le1/9),\qquad \beta(s)=\frac{324s^2-4}{5}\quad(1/9<s\le \delta).
\]

It joins zero to one continuously. Set
\[
\zeta=(1-s)(1-2\theta)/2+\beta Z.
\]
By differentiation in \eqref{eq:stream}, the radial part gives
$\phi(s)h/s$. In the reserve problem, the correction in
\cref{sec:reserve-field} replaces this part by $\phi(s)\one/2$, leaving
the stream correction and the component sum unchanged.
Its contribution on winner levels is nonnegative by
\cref{sec:reserve-proof}. The streams match at both junctions
and the stream correction vanishes on winner axes.
On a tie, the extra normal sum is $(\beta/s)Z_\theta$, which is
reciprocal because $Z_\theta=-\frac{11}{24}(2\theta-1)(2\eta-1)$ there.
Thus all required normal traces hold.

Differentiation of the factored stream gives the first component of the correction on $1/9<s\le \delta$:

\[
-\frac{11}{24}\theta\left[\frac\beta s\theta+\frac8{5s}(1-\theta)\right](\eta-\nu).
\]

This has nonpositive derivative in $l_1$, and the mirror has nonpositive derivative in $l_2$. Consequently the loser volume deficit is nonnegative. For the exact radial-based field, nonpositivity of both components follows from

\[
\phi(s)+\frac{11\beta}{24s}\le\phi(\delta)+\frac{11}{4}=-3,
\]

\[
\phi(s)+\frac{11}{15s}=\frac32s-\frac4{15s}<0.
\]

For the equal-component reserve field use $\beta/s\le6$ and $8/(5s)\le72/5$. The correction's absolute value is at most

\[
\frac{11}{24}\max_{0\le\theta\le1}\left[6\theta^2+\frac{72}{5}\theta(1-\theta)\right]=\frac{99}{35}<\frac{23}{8}=-\frac{\phi(\delta)}2.
\]

Hence both fields are componentwise nonpositive. Below $1/9$ this is immediate. The loser residual on a low triangle is therefore a nonnegative IR measure, with no row calculation needed.

For the reserve floor on $[1/9,\delta]$, the negative of the reserve field's first component is bounded above by

\[
\frac1{2s}-\frac{3s}4+\frac{11}{24}\theta\left[\frac\beta s\theta+\frac8{5s}(1-\theta)\right].
\]

For fixed $\theta$ this is affine in $s$ plus a nonnegative
multiple of $1/s$: the coefficient of $1/s$ is
$1/2+\frac{11}{120}(8\theta-12\theta^2)\ge2/15$.
At the lower endpoint use the trace of the upper branch of $\beta$. Thus it is convex in $s$ and its maximum is at an endpoint. At $s=1/9$ it is at most

\[
\frac{53}{12}+\frac{33}{20}=\frac{91}{15}.
\]

At $s=\delta$ it is at most $23/8+11/4=45/8$ (since $\beta/s=6$ and $8/(5s)=48/5\le12$). This proves the floor on the bridge.

For $c\ge91/15$, the reserve equation implies $\varepsilon<1/(2c)<1/9$. Below $1/9$ the sold reserve field is $\phi(s)\one/2$; its components are at least $-c$ since $\phi$ increases and $\phi(\varepsilon)=-2c$. The no-sale field is $-c(1,1)$; its normal sum matches the exact winner field at the reserve boundary. Its IR residual is nonnegative. No singularity is present in the actual reserve field.

\section{Upper region}\label{sec:upper-verification}
\subsection{Upper strip}

When $\one\cdot h\ge1$, the relevant winner boundary includes the top
and right edges of the type square. Complementary coordinates
turn these into zero-coordinate faces and allow us to reuse
the same kernel and boundary profile. Use complementary types
$H=(1,1)-h$, $L=(1,1)-l$, with $\sigma=H_1+H_2$,
$\rho=L_1+L_2$ and $\sigma\le\rho$. Write the upper score as
$S_0=(1,1)-T$. For $\delta\le\sigma\le1$ use

\[
\begin{aligned}
\zeta={}&H_2P(L_1-H_1)-H_1P(L_2-H_2)\\
&-\sigma[P(L_1)G(H_1)-P(L_2)G(H_2)],
\end{aligned}
\]

\[
T=\frac32H+(-\partial_{H_2}\zeta,\partial_{H_1}\zeta).
\]

This is exactly the middle strip's stream at $\one\cdot h=1$. On $H_j=0$ it has zero trace. On $H_1=L_1+d$ it equals $H_2/2$ by saturation, giving the required split normal data; the other face is mirrored. On ties its angular derivative is reciprocal by the same even-kernel/product argument as in the middle strip.

Differentiating, with $e_j=L_j-H_j$, yields

\[
\begin{aligned}
T_1={}&3H_1/2-P(e_1)+H_1k(e_2)\\
&+P(L_1)G(H_1)-P(L_2)[G(H_2)-48\sigma P(H_2)].
\end{aligned}
\]

The bracket $G(t)-48\sigma P(t)$ is nonpositive, since $\sigma\ge \delta$ and $G(t)\le 2(1-6t)_+\le48\sigma P(t)$. The pure derivative in $L_1$ is $k(e_1)-k(L_1)G(H_1)\le3/2$. The tie slack is

\[
\psi(e_1)+H_1k(e_2)+P(L_1)G(H_1)-P(L_2)[G(H_2)-48\sigma P(H_2)]\ge0.
\]

At a physical loser edge $L_1=0$, write $x=H_1\le d$. Since $L_2\ge\sigma\ge \delta$, the last term vanishes, and $e_2\ge x$. Thus

\[
T_1\le3x/2-P(-x)+xk(x)+G(x)/4.
\]

For $0\le x<\delta$, the right side is $(3x/2)(6x-1)\le0$; for $\delta<x\le d$ it is $3x/2-1/2\le0$. This proves the physical-edge condition, including relevant one-sided traces. Split endpoints have the usual slack. All row conditions hold. Finally $T_1\le3+1/4+3=25/4$ by $k\le3/2$, $P\le1/4$ on nonnegative inputs, $G\le1$, $\sigma\le1$, and $48P(t)\le12$ for $t\ge0$. Hence $S_{0,1}\ge-21/4>-91/15$. The mirror gives all item-2 statements. There is no reserve correction in this region.

\subsection{Upper cap}\label{sec:top-hom}

Near the upper corner there are no split faces. For losers with
$\rho\ge\delta$, we extend the stream at $\sigma=\delta$ using
truncated linear terms that preserve the zero-axis traces and
control the loser-side slopes. For $0<\sigma\le\delta$ and
$\rho\ge\delta$, define

\[
\begin{aligned}
\zeta=\frac32\bigg[&H_1H_2\frac{\min(L_2,\delta)-\min(L_1,\delta)}\delta\\
&-H_2\min(H_1,(L_1-\delta)_+)\\
&+H_1\min(H_2,(L_2-\delta)_+)\bigg].
\end{aligned}
\]

At $\sigma=\delta$, the two stream formulas agree. Indeed,
$H_1+H_2=\delta$ and the profile formulas give
\[
\begin{aligned}
&H_2P(L_1-H_1)-\delta P(L_1)G(H_1)\\
&=\frac32\bigg[
 H_1H_2\left(2-\frac{\min(L_1,\delta)}{\delta}\right)
 -H_2\min(H_1,(L_1-\delta)_+)\bigg].
\end{aligned}
\]
For $L_1\le\delta$ this follows from the linear profile and
$G(H_1)=(H_2/\delta)^2$. For $L_1\ge\delta$ the second
term on the left vanishes, and
$P(L_1-H_1)=\frac32(H_1+\delta-L_1)_+$.
Subtracting the item-swapped identity gives the displayed
cap stream.

When the loser also lies near the upper corner, we can use a
simpler product field: its dependence on the two types makes
the tie sum reciprocal and gives a direct bound on the total
loser divergence. For $0<\rho<\delta$ use instead

\[
\zeta=\frac32 H_1H_2\frac{L_2-L_1}{\rho},
\]

which gives the field

\[
T_j=\frac{3H_jL_j}{\rho}\quad(j=1,2).
\]

Both streams vanish on the winner axes. There are no split faces for $\sigma<\delta<d$. The fields match at $\rho=\delta$, since then both $L_j\le \delta$ and the min corrections vanish. Tie reciprocity in the inner block is immediate from the symmetric product on $\sigma=\rho$. At $H=L=0$ take $T=0$; exceptional values do not affect volume integrals. Winner divergence is 3 by the stream formula (or direct differentiation in the product block).

In the outer cap $\rho\ge \delta$, differentiation gives

\[
\begin{aligned}
T_1={}&\frac32H_1\left[1-\frac{\min(L_2,\delta)-\min(L_1,\delta)}\delta\right]\\
&+\frac32\min(H_1,(L_1-\delta)_+)\\
&-\frac32H_1\,\mathbf1_{\{H_2<(L_2-\delta)_+\}}.
\end{aligned}
\]

As a function of $L_1$, it is continuous and its slope is $3H_1/(2\delta)$ below $\delta$, $3/2$ on the active min interval above $\delta$, and zero otherwise. All these slopes are at most $3/2$. At $L_1=0$ the first two lines vanish, and the last line is nonpositive. In the inner block

\[
\dv_L T=\frac{3(H_1L_2+H_2L_1)}{\rho^2}\le3,
\]

because $H_1+H_2\le\rho$. Also $T_j=0$ when $L_j=0$. At entry $\rho=\delta$ the product formula gives $T_j\ge0$, and the partial row slack is

\[
3l_j/2-S_{0,j}=1/2-3L_j/2+T_j\ge1/4.
\]

For a fixed winner in the cap, any row either ends at a physical edge outside the inner block, or enters the inner block at $\rho=\delta$ before its endpoint. The former is paid by the pure deficit and physical-edge condition, and the latter by the entry slack followed by half the nonnegative total deficit in the block. This is the row argument of \cref{prop:H}. There is no additional interface atom at $\rho=\delta$. The column proof is identical.

In the outer cap $T_j\le(9/2)H_j\le3/4$; in the inner cap $T_j\le3H_j\le1/2$. These imply the floor with room to spare.

\subsection{Regularity and completion}\label{sec:outer-completion}
All streams match across winner-level junctions. On almost every
fixed loser slice, they are continuous and piecewise smooth with
bounded first derivatives on a finite partition of the winner cell.
The lower winner total is bounded below by the positive loser total;
in the inner upper cap $\rho>0$ is fixed in the winner variable.
Thus the weak divergence and matching normal fluxes follow from
\eqref{eq:stream}. All tie, split and physical boundary conditions
refer to one-sided traces; kink intersections on these faces are
null for almost every opponent. At the upper-to-middle junction the
complementary coordinate transformation changes both the rotated
gradient and the sign of the field, so the two formulas use the
same stream.

On almost every fixed winner slice the loser fields are piecewise
smooth with bounded derivatives on finitely many pieces; in the
inner cap $\rho\ge\sigma>0$. The middle own-coordinate jumps are
downward. In the upper strip and outer cap each component's jumps
depend only on the other loser coordinate, so they add no divergence
measure. At $\rho=\delta$ both cap streams agree for all eligible
$H$; differentiating in the winner variables shows that the fields
match there almost everywhere. Hence the distributional divergence
agrees with the row and column Stieltjes derivatives by Fubini.
This justifies the integration by parts and the row criterion in
\cref{sec:propH}. The only unbounded exact-field behavior across
profiles is $O(1/(\one\cdot h))$ near the radial origin. It is integrable
against the lower-total density $\one\cdot h$; the actual reserve field
is bounded since its sold levels satisfy $\one\cdot h>\eps>0$.
Consequently all conditional integrations and expectations exist.

\subsection{Completion of the field construction}\label{sec:threshold}

\begin{proof}[Proof of \cref{lem:dual-fields}]
The preceding constructions define the common fields throughout
the bundle regions. Their measurability, integrability, and
normal traces follow from \cref{sec:outer-completion}.
The loser and no-sale residuals derived in \cref{sec:propH}
are nonnegative by the regional verifications and vanish at
the candidate, establishing property (i).
The fixed-total correction in \cref{sec:reserve-field,sec:middle-correction}
has the support, zero integral, and sign pattern of property (ii).
Finally, the middle, lower and upper bounds give
\[
S_{\eps,j}(h,l)\ge-c,\qquad \one\cdot h>\eps,\quad c\ge91/15,\quad j=1,2
\]
almost everywhere on sold bundle pairs, proving property (iii).
\end{proof}


\nocite{*}
\printbibliography[title={References}]

\end{document}